\documentstyle[preprint,aps]{revtex}
\begin{document}
\setlength{\baselineskip}{18pt}
\begin{flushright}
TU-TP/98-01\\
January, 1998
\end{flushright}
\begin{center}
{\Large \bf Analysis of neutrino oscillation in three-flavor neutrinos}\\
\vspace{1cm}
T. Teshima\thanks{E-mail address: teshima@isc.chubu.ac.jp}
, T. Sakai and O. Inagaki\\
\vspace{1cm}
Department of Applied Physics,  Chubu University, Kasugai 487, Japan\\
\end{center}
\begin{abstract}
We analyzed the solar, terrestrial and atmospheric neutrinos experiments using the three-flavor neutrino framework and got the allowed regions for parameters $(\Delta m_{12}^2,\ \sin^22\theta_{12},\ \Delta m_{23}^2,\ \theta_{13},\ \theta_{23})$. In solar neutrino experiments, we got the large angle solution $(\Delta m_{12}^2,\ \sin^22\theta_{12})=(4\times10^{-6} - 7\times10^{-5}{\rm eV}^2,\ 0.6 - 0.9)$ and small angle solution $(3\times10^{-6} - 1.2\times10^{-5}{\rm eV}^2,\ 0.003 - 0.01)$ for $\theta_{13}=0^\circ - 20^\circ$. From the terrestrial and atmospheric neutrino experiments, we got the allowed regions $(\theta_{13}<4^\circ,\ 24^\circ<\theta_{23}<26^\circ)$ for $\Delta m_{23}^2=2{\rm eV}^2$, $(\theta_{13}<4^\circ,\ 24^\circ<\theta_{23}<45^\circ)$ for $\Delta m_{23}^2=0.2{\rm eV}^2$ and $(\theta_{13}<14^\circ,\ \theta_{23}\sim40^\circ)$ for $\Delta m_{23}^2=0.02{\rm eV}^2$. It seems that the large angle solution for $(\Delta m_{12}^2,\ \sin^22\theta_{12})$ is favored than the small angle slution from the analysis of zenith angle dependence in atmospheric neutrino sub-GeV experiment. 
\end{abstract}
\vspace{0.5cm}
\setlength{\baselineskip}{16pt}
\section{Introduction}
The problem of neutrino masses and oscillations is one of the most interesting issues to study physics beyond the standard model (SM) \cite{STANDARD}. In many experiments which are under way, indications in favor of neutrino masses and oscillations have been obtained. In these experiments, the solar neutrino experiments \cite{SAGE,GALLEX,HOMESTAKE,KAMIOKANDE} measure the event rates significantly lower than the ones predicted by the standard solar model, and the atmospheric neutrino experiments \cite{HIRATA,FUKUDA,IMB} observe an anomaly referred to as the unexpected difference between the measured and predicted $\mu$-like and $e$-like neutrinos. Another indication in favor of non-zero neutrino masses is in the cosmological analysis by dark matter \cite{SCHRAMM}. 
\par 
On the other hand, terrestrial neutrino experiments searching for the neutrino masses and oscillations are under way. The E531 \cite{E531}, CHORUS and NOMAD \cite{CHORUS} experiments using the beam from accelerator search for $\nu_{\tau}$ appearance in a $\nu_{\mu}$, and E776 \cite{E776}, KARMEN \cite{EMU} and LSND \cite{LSND} experiments using the accelerator beams are searching for $\nu_{\mu} \to \nu_e$ and $\bar{\nu}_{\mu} \to \bar{\nu}_e$ oscillations. The experiments using nuclear power reactor \cite{ENOTE} search for the disappearance of $\bar{\nu}_e$, in which $\bar{\nu}_e \to \bar{\nu}_X\ (X=\mu,\,\tau)$ transitions are expected. These experiments do not observe significantly large neutrino transitions.
\par
In this paper, we analyze the data of current solar, terrestrial and atmospheric experiments in a framework where the three neutrinos have masses and mix each other, and search the allowed regions of parameters $(\Delta m_{12}^2,\ \sin^22\theta_{12},\ \Delta m_{23}^2,\ \theta_{13},\ \theta_{23})$ characterizing the masses and mixing of three-flavor neutrinos. Although there are many analyses which study the solar, atmospheric and terrestrial neutrino problems in three-flavor neutrinos framework \cite{MIKHEYEV,KUO,FOGLI}, we study more thoroughly these problems including the analysis of zenith angle dependence in recent SuperKamiokande atmospheric experiment \cite{SUPERKAMIOKANDE,KANEYUKI}. After the analyses of experiments for neutrino oscillation, we will present a matrix of neutrino mixing ascribed from the allowed regions of parameters obtained. 

\section{Neutrino oscillation}
Weak currents for the interactions producing and absorbing neutrinos are described as 
\begin{equation}
J_{\mu}=2\sum^3_{\alpha,\beta=1}\bar{l}_{L\alpha}\gamma_{\mu}U_{l_{\alpha}\beta}\nu_{L\beta},
\end{equation}
where $l_{\alpha}\ (l_1=e,\ l_2=\mu,\ l_3=\tau)$ represents the lepton flavor, $\nu_{\beta}$ the neutrino mass eigenstate and $U$ is the lepton mixing matrix.
$U$ is the unitary matrix corresponding to the CKM matrix $V_{CKM}^{\dagger}$ for quarks defined by
\begin{equation}
U=U_lU^{\dagger}_{\nu},
\end{equation}
where the unitary matrices $U_l$ and $U_{\nu}$ transform mass matrices $M_l$ for charged leptons and $M_{\nu}$ for neutrinos to diagonal mass matrices as
\begin{equation}
\begin{array}{l}
U_lM^lU^{-1}_l={\rm diag}[m_e, m_{\mu}, m_{\tau}], \\
U_{\nu}M^{\nu}U^{-1}_{\nu}={\rm diag}[m_1, m_2, m_3].
\end{array}
\end{equation}
We present the unitary matrix neglecting the {\it CP} violation phases as 
\begin{eqnarray}
 U&=&e^{i\theta_{23}\lambda_7}e^{i\theta_{13}\lambda_5}e^{i\theta_{12}
             \lambda_2}\nonumber\\
            &=&\left(
      \begin{array}{ccc}
      c_{12}^{\nu}c_{13}^{\nu} & s_{12}^{\nu}c_{13}^{\nu} & s_{13}^{\nu} \\
      -s_{12}^{\nu}c_{23}^{\nu}-c_{12}^{\nu}s_{23}^{\nu}s_{13}^{\nu} & c_{12}^{\nu}c_{23}^{\nu}-s_{12}^{\nu}s_{23}^{\nu}s_{13}^{\nu} & s_{23}^{\nu}c_{13}^{\nu} \\
      s_{12}^{\nu}s_{23}^{\nu}-c_{12}^{\nu}c_{23}^{\nu}s_{13}^{\nu} & -c_{12}^{\nu}s_{23}^{\nu}-s_{12}^{\nu}c_{23}^{\nu}s_{13}^{\nu} & c_{23}^{\nu}c_{13}^{\nu} 
      \end{array}\right), \\
 & & \quad\quad c_{ij}^{\nu}=\cos{\theta}^{\nu}_{ij},\ \  s_{ij}^{\nu}=\sin{\theta}^{\nu}_{ij}, \nonumber 
\end{eqnarray}  
where $\lambda_i$'s are Gell-Mann matrices.
\par 
The probabilities for transitions  $\nu_{l_\alpha} \to \nu_{l_\beta}$ are written as 
\begin{eqnarray}
P(\nu_{l_\alpha}\to\nu_{l_\beta})&&=|<\nu_{l_\beta}(t)|\nu_{l_\alpha}(0)>|^2 = \delta_{l_\alpha l_\beta}+p_{\nu_{l_\alpha}\to\nu_{l_\beta}}^{12}S_{12}+p_{\nu_{l_\alpha}\to\nu_{l_\beta}}^{23}S_{23}+p_{\nu_{l_\alpha}\to\nu_{l_\beta}}^{31}S_{31},\nonumber \\
&&p_{\nu_{l_\alpha}\to\nu_{l_\beta}}^{12}=-2\delta_{l_\alpha l_\beta}(1-2U_{l_\alpha3}^2)+2(U_{l_\alpha1}^2U_{l_\beta1}^2+U_{l_\alpha2}^2U_{l_\beta2}^2-U_{l_\alpha3}^2U_{l_\beta3}^2), \nonumber \\       
&&p_{\nu_{l_\alpha}\to\nu_{l_\beta}}^{23}=-2\delta_{l_\alpha l_\beta}(1-2U_{l_\alpha1}^2)+2(-U_{l_\alpha1}^2U_{l_\beta1}^2+U_{l_\alpha2}^2U_{l_\beta2}^2+U_{l_\alpha3}^2U_{l_\beta3}^2), \nonumber \\
&&p_{\nu_{l_\alpha}\to\nu_{l_\beta}}^{31}=-2\delta_{l_\alpha l_\beta}(1-2U_{l_\alpha2}^2)+2(U_{l_\alpha1}^2U_{l_\beta1}^2-U_{l_\alpha2}^2U_{l_\beta2}^2+U_{l_\alpha3}^2U_{l_\beta3}^2) ,\nonumber \\
&&
\end{eqnarray}
where $S_{ij}$ is the term representing the neutrino oscillation;  
\begin{equation}
S_{ij}=\sin^21.27\frac{\Delta m^2_{ij}}{E}L,
\end{equation}
in which $\Delta m^2_{ij}=|m^2_i-m^2_j|$, $E$ and $L$  are measured in units eV$^2$, MeV and m, respectively. From the unitarity of $U$, we get relations 
\begin{equation}
p^{ij}_{\nu_{l_\alpha}\to\nu_e}+p^{ij}_{\nu_{l_\alpha}\to\nu_{\mu}}+p^{ij}_{\nu_{l_\alpha}\to\nu_{\tau}}=0, \ \ \ \ i,\ j=1, 2, 3, \ \ \ \ l_\alpha=e, \mu, \tau.
\end{equation}
\par
The values of neutrino masses are not known precisely, but we know that if one identifies the dark matter of universe (or at least its hot dark matter component) with neutrino matter one has \cite{SCHRAMM} 
\begin{equation}
m_1+m_2+m_3 \sim {\rm several \ eV}. 
\end{equation}
We do not use this value strictly in present analysis, but we consider that the sum of neutrino masses is not so small. In two-flavor neutrinos analyses in which one mass parameter $\Delta m^2$ appears for solar neutrino experiments, the result that $\Delta m^2$ is $10^{-4} - 10^{-5}{\rm eV}^2$ or $\sim 10^{-10}{\rm eV}^2$ is obtained \cite{KAMIOKANDE3}.  For atmospheric experiments, $\Delta m^2$ is obtained as $10^{-1} - 10^{-2}{\rm eV}^2$ \cite{KAMIOKANDE3}. Then it seems that two neutrinos masses in three neutrinos are very close and another one is rather far away from them. Then we assume that three neutrino masses have such a mass hierarchy as  
\begin{equation}
m_1 \approx m_2 \ll m_3. \label{masshie}
\end{equation}
In the the mass hierarchy Eq.~(\ref{masshie}), $\Delta m_{12}^2\ll\Delta m_{23}^2 \simeq \Delta m_{13}^2$, the expression Eq. (5) for the transition probabilities $P(\nu_{l_{\alpha}}\to\nu_{l_{\beta}})$ are rewritten as
\begin{eqnarray}
P(\nu_e\to\nu_e)&=&1-2(1-2U_{e3}^2-U_{e1}^4-U_{e2}^4+U_{e3}^4)S_{12}-4U_{e3}^2(1-U_{e3}^2)S_{23}, \nonumber \\
P(\nu_{\mu}\to\nu_{\mu})&=&1-2(1-2U_{\mu3}^2-U_{\mu1}^4-U_{\mu2}^4+U_{\mu3}^4)S_{12}-4U_{\mu3}^2(1-U_{\mu3}^2)S_{23}, \nonumber \\
P(\nu_\tau\to\nu_\tau)&=&1-2(1-2U_{\tau3}^2-U_{\tau1}^4-U_{\tau2}^4+U_{\tau3}^4)S_{12}-4U_{\tau3}^2(1-U_{\tau3}^2)S_{23}, \nonumber \\   
P(\nu_\mu\to\nu_e)&=&P(\nu_e\to\nu_\mu)=2(U_{\mu1}^2U_{e1}^2+U_{\mu2}^2U_{e2}^2-U_{\mu3}^2U_{e3}^2)S_{12}+4U_{e3}^2U_{\mu3}^2S_{23}, \nonumber \\
P(\nu_\tau\to\nu_e)&=&P(\nu_e\to\nu_\tau)=2(U_{\tau1}^2U_{e1}^2+U_{\tau2}^2U_{e2}^2-U_{\tau3}^2U_{e3}^2)S_{12}+4U_{e3}^2U_{\tau3}^2S_{23}, \nonumber \\
P(\nu_\tau\to\nu_\mu)&=&P(\nu_\mu\to\nu_\tau)=2(U_{\tau1}^2U_{\mu1}^2+U_{\tau2}^2U_{\mu2}^2-U_{\tau3}^2U_{\mu3}^2)S_{12}+4U_{\mu3}^2U_{\tau3}^2S_{23}.\nonumber \\
&& 
\end{eqnarray}

\section{Numerical analyses of neutrino oscillation in three-flavor neutrinos}
\subsection{Solar neutrinos}
We first analyze the solar neutrino experiments. Considering the matter effects (MSW effect \cite{MSW}) in three-flavor neutrinos, the transition probability for $\nu_e \to \nu_e$ is expressed as \cite{KUO}
\begin{mathletters}
\begin{eqnarray}
&&P^{\rm MSW}_{3\nu}(\Delta m_{12}^2,\,\theta_{12},\,\theta_{13},\,E)=\cos^4\theta_{13}P^{\rm MSW}_{2\nu}(\Delta m_{12}^2,\,\theta_{12},\,\theta_{13},\,E)+\sin^4\theta_{13}C\\ 
&&P^{\rm MSW}_{2\nu}(\Delta m_{12}^2,\,\theta_{12},\,\theta_{13},\,E)=\frac12+\left(\frac12-\Theta(A\cos^2{\theta_{13}}-\Delta m_{12}^2\cos{2\theta_{12}})P_c(\theta_{12}, E)\right)\nonumber \\
&&\qquad\qquad\qquad\times\cos2\theta_{12}\cos2\theta_{12}^m,\\
&&P_c(\theta_{12},\,E)=\exp{(-\frac{\pi}{2}\gamma(\theta_{12}, E))},\\
&&\gamma(\theta_{12},\,E)=\frac{\Delta m_{12}^2\sin^2{2\theta_{12}}}{2E\cos{2\theta_{12}}|dN_e/N_edx|_0},
\end{eqnarray}
\end{mathletters}
where $\Theta$ is the theta function, $A=2\sqrt{2}G_FN_eE$ ($G_F$ is Fermi constant, $N_e$ number of electron per cm$^3$ at the production point of neutrinos in the sun and $E$  the energy of neutrino), $P_c$ the Landau-Zener-Stueckerberg crossing probability, $\gamma$ is the adiabaticity parameter ($|\cdots|_0$ represents the value at the production point of neutrinos) and $\theta^m_{12}$ the mixing angle at the production point. $P^{\rm MSW}_{2\nu}$ is the transition probability with a replacement of $N_e \to N_e\cos2\theta_{13}$ in two-flavor neutrinos transition probability. This expression is obtained from an approximation; 
\begin{equation}
\Theta[A-(m_3^2-\Lambda/2)\cos2\theta_{13}]\exp\left(-\frac{\pi}{4}\frac{(m^2_3-\Lambda/2)}{|dN_e/N_edx|_0E}\frac{\sin^22\theta_{13}}{\cos2\theta_{13}}\right)\ll1,
\end{equation}
where $\Lambda=((m^2_1+m^2_2)-(m^2_2-m^2_1)\cos2\theta_{12})$. This approximation is reasonable for present assumption of mass hierarchy Eq.~(9) because of $A\ll m^2_3-\Lambda/2$ and $m^2_3-\Lambda/2\gg|dN_e/N_edx|_0E$.
\par
The ratios of the detected $e$ neutrino fluxes to the expected $e$ neutrino fluxes deduced from the standard solar model (SSM)\cite{BAHCALL} are expressed by using the transition probability $P^{\rm MSW}_{3\nu}$ as
\begin{equation}
R=\frac{\int^{E_{max}}_{E_{min}} P^{\rm MSW}_{3\nu}(E)f(E)dE}{\int^{E_{max}}_{E_{min}} f(E) dE},
\end{equation}
where $f(E)$ is the product of the spectral function of neutrino flux and detector sensitivity. The neutrino flux is the sum of many neutrino fluxes produced by the various nuclear fusion reaction at the center of the sun. The detector sensitivity also depends to the neutrino energy. Numerical results for neutrino flux and detector sensitivity are analyzed precisely in Ref. \cite{BAHCALL}. For $f(E)$ used in this paper, see Appendix A.
\par
In our analysis, we use the following three experimental data for $R$:\\
\begin{mathletters}
\hspace*{1cm} Ga experiment\cite{SAGE,GALLEX}:
\begin{equation}
R=0.534\pm0.087,
\end{equation}
\hspace*{1cm} Cl experiment\cite{HOMESTAKE}:
\begin{equation}
R=0.274\pm0.046,
\end{equation}
\hspace*{1cm} water Cherenkov experiment\cite{KAMIOKANDE}:
\begin{equation} 
R=0.437\pm0.092.
\end{equation}
\end{mathletters}
In Ga experiment, we combined the SAGE\cite{SAGE} and GALLEX\cite{GALLEX} data. First we estimate the numerical values $R$ for $\theta_{13}=0$ which corresponds to two-flavor neutrinos case using the Eq.~(13), and show the contours of $R$ on $\sin^22\theta_{12}-\Delta m_{12}^2$ plane in Fig.~\ref{fig1}.  Each contour denoted as Ga, Cl and Kam corresponds to the upper and lower values of $R$ in Eq.~(14) for Ga, Cl and Kamiokande's water Cherenkov experiment.  There are two solutions which are denoted as common areas enclosed by each two contours of Ga, Cl and Kam. These are called as large solution and small solution. This result is similar to the one obtained in various analyses \cite{HATA}. 
\par
Next we estimate the numerical values $R$ for the case of $\theta_{13}\ne0$. In Fig.~\ref{fig2}, we show the allowed regions of the combined Ga, Cl and Kam experiments using the $\chi$-square, where the solid thin, solid thick and  dotted lines define the regions allowed at $99 \%(\chi^2=9.2)$, $95 \%(\chi^2=6.0)$ and $90 \%(\chi^2=4.5)$ C.L., respectively. Fig.~\ref{fig2}(a) -  Fig.~\ref{fig2}(h) show the allowed regions for $\theta_{13}=0^{\circ} - 50^{\circ}$. Fig.~\ref{fig2}(a) shows the $\theta_{13}=0$ case, thus this shows the two-flavor neutrinos' solution. These results are similar to the ones obtained by Ref.~\cite{FOGLI}. Numerically, we show the allowed regions in $95 \%$ C.L.;  
\begin{mathletters}
\begin{eqnarray}
{\rm for}&&\ \  \theta_{13}=0^\circ-20^\circ \nonumber\\
&&\left.\begin{array}{l}
\Delta m_{12}^2=4\times10^{-6}\ - 7\times10^{-5} {\rm eV}^2 \\
\sin^22\theta_{12}=0.6-0.9      
\end{array} \right\}{\rm large\ angle\ solution}\\
&&\left.\begin{array}{l}
\Delta m_{12}^2=3\times10^{-6}\ - 1.2\times10^{-5} {\rm eV}^2 \\
\sin^22\theta_{12}=0.003-0.01      
\end{array} \right\} {\rm small\ angle\ solution}\\
{\rm for}&&\ \  \theta_{13}=25^\circ-40^\circ \nonumber\\
&&\left.\begin{array}{l}
\Delta m_{12}^2=2\times10^{-6}\ - 3\times10^{-5} {\rm eV}^2 \\
\sin^22\theta_{12}=0.001-0.01      
\end{array} \right\} {\rm small\ angle\ solution}\\
{\rm for}&&\ \  \theta_{13}=45^\circ-50^\circ \nonumber\\
&&\left.\begin{array}{l}
\Delta m_{12}^2=2\times10^{-6}\ - 3\times10^{-5} {\rm eV}^2 \\
\sin^22\theta_{12}=0.001-0.7      
\end{array} \right\} 
\end{eqnarray}
\end{mathletters}\\
The characteristic feature of three-flavor neutrinos' solution is as follows; increasing $\theta_{13}$ from $0^{\circ}$ to $25^{\circ}$, the small mixing solution and the large mixing solution merge in a single solution, further increasing $\theta_{13}$ to $50^{\circ}$, the allowed region becomes broader and next shrinks and lastly disappears.       

\subsection{Terrestrial neutrinos}
In terrestrial experiments, there are two types of experiment: short baseline and long baseline experiment. In present study, we analyze the short baseline experiment. In the short baseline experiments, there are E531 \cite{E531}, CHORUS and NOMAD \cite{CHORUS} accelerator experiments searching for $\nu_{\tau}$ appearance in $\nu_{\mu}$. We use the data of E531, CHORUS and NOMAD experiments;
\begin{eqnarray}
&&P(\nu_{\mu}\to\nu_{\tau})<2\times10^{-3}\ \ (90\%\ {\rm C.L.}),\\
&&\qquad\qquad L/E\sim0.02.\nonumber
\end{eqnarray}
For the experiments searching for $\nu_{\mu}\to\nu_{e}$ and $\bar{\nu}_{\mu}\to\bar{\nu}_e$ oscillations, there are  E776 \CITE{E776}, KARMEN \cite{EMU} and LSND \cite{LSND} accelerator experiments;
\begin{mathletters}
\begin{eqnarray}
&&P(\nu_{\mu}\to\nu_{e})<3\times10^{-3}\ \ (90\%\ {\rm C.L.}),\ \ {\rm E776}\\
&&\qquad\qquad L=1{\rm km},\ \ \ \ E \sim 1{\rm GeV},\nonumber\\
&&P(\bar{\nu}_{\mu}\to\bar{\nu}_{e})<3.1\times10^{-3}\ \ (90\%\ {\rm C.L.}),\ \ {\rm KARMEN}\\
&&\qquad\qquad L=17.5{\rm m},\ \ \ \ E<50{\rm MeV},\nonumber\\
&&P(\bar{\nu}_{\mu}\to\bar{\nu}_{e})=3.4\mbox{\small${+2.0\atop-1.8}$}\pm0.7\times10^{-3},\ \  {\rm LSND}\\
&&\qquad\qquad  L=30{\rm m},\ \ \ \ E\sim36-60{\rm MeV}.\nonumber
\end{eqnarray}
\end{mathletters}
 Furthermore, we analyse the experiments using nuclear power reactor \cite{ENOTE} searching for the disappearance of $\bar{\nu}_e$,
\begin{eqnarray}
&&1-P(\bar{\nu}_{e}\to\bar{\nu}_{e})<10^{-2}\ \ (90\%\ {\rm C.L.}),\\
&&\qquad\qquad L=15,\ 40,\ 95{\rm m},\ \ \ E\sim1-6{\rm MeV}.\nonumber
\end{eqnarray}
\par In the short baseline experiments, the neutrino propagation length $L$ is about $\displaystyle{L\stackrel{<}{\sim}1{\rm km}}$, then $\displaystyle{S_{12}=\sin^21.27\frac{\Delta m^2_{12}}{E}L}$ is very small because $\Delta m^2_{12} \sim 10^{-5} - 10^{-4}{\rm eV}^2$ suggested from last solar neutrino analyses. Then, the $S_{23}$ term is dominant in the transition probability Eq.~(10). Seeing that the mixing parts proportional to $S_{23}$ are $4U^2_{e3}(1-U^2_{e3})$, $4U^2_{\mu3}(1-U^2_{\mu3})$ and $4U^2_{\tau3}(1-U^2_{\tau3})$, the transition probabilities in present short baseline terrestrial experiments seem to be insensitive to the parameters; $\Delta m^2_{12}$ and $\theta_{12}$. 
\par 
We show the contour plots of the allowed regions on $(\tan^2\theta_{13},\ \tan^2\theta_{23}$) plane determined by the probability $P$ expressed in Eq.~(10) and above experimental data Eqs.~(16), (17) and (18) in Fig.~\ref{fig3}. Allowed regions are corners, left and right hand sides surrounded by curves. Curves represent the boundary of $90$ \% C.L. of $P$. We fixed the parameters $\Delta m_{12}^2$ and $\theta_{12}$ as $\Delta m_{12}^2=10^{-5}{\rm eV}^2$ and $\sin^22\theta_{12}=0.8$, and the parameter $\Delta m_{13}^2$ to be various values from 0.02eV$^2$ to 20eV$^2$. Although we fix the parameters $\Delta m_{12}^2$ and $\theta_{12}$ as $\Delta m_{12}^2=10^{-5}{\rm eV}^2$ and \ $\sin^22\theta_{12}=0.01$, the results are not changed. Dotted lines show the allowed regions restricted by the LSND data.  From these results, we see that the numerical allowed regions on $(\theta_{13},\ \theta_{23})$ without the LSND data are as follows;
\begin{mathletters}
\begin{eqnarray}
{\rm for}\  \ && \Delta m_{23}^2=20{\rm eV}^2\nonumber\\
&&(<4^\circ,\ <2^\circ),\ \ (<2^\circ,\ >88^\circ),\ \ (>86^\circ-88^\circ,\ {\rm any}),\\ 
{\rm for}\ \ && \Delta m_{23}^2=2\,{\rm eV}^2\nonumber\\
&&(<4^\circ,\ <26^\circ),\ \ (<2^\circ,\ >65^\circ),\ \ (>86^\circ-88^\circ,\ {\rm any}),\\
{\rm for}\ \ &&\Delta m_{23}^2=0.2\,{\rm eV}^2\nonumber\\  
&&(<4^\circ,\ {\rm any}),\ \  (>86^\circ,\ {\rm any}),\\
{\rm for}\ \ &&\Delta m_{23}^2=0.02\,{\rm eV}^2\nonumber\\ 
&&(<12^\circ,\ {\rm any}),\ \ (>78^\circ,\ {\rm any})\\ 
{\rm for}\ \ &&\Delta m_{23}^2<0.005{\rm eV}^2\nonumber\\
&&{\rm all\ regions\ are\ allowed.}
\end{eqnarray}
\end{mathletters}  
\par
If we combine the LSND data with the above analyses, the allowed region disappears lower than 0.2\,eV$^2$ of the $\Delta m_{23}^2$ value. Furthermore, combining the solar neutrino solutions with this terrestrial ones, the allowed regions larger than 50$^\circ$ of $\theta_{13}$ on terrestrial neutrino are excluded.

\subsection{Atmospheric neutrinos}
\par
The evidence for an anomaly in atmospheric neutrino experiments was pointed out  by the Kamiokande Collaboration \cite{HIRATA,FUKUDA} and IMB Collaboration \cite{IMB} using the water-Cherencov experiments. More recently, SuperKamiokande Collaboration \cite{SUPERKAMIOKANDE,KANEYUKI} reports the more precise results on anomaly in atmospheric neutrino. We tabulate these results in Table \ref{table2}. That the double ratio, $R(\mu/e)\equiv{R_{\rm expt}(\mu/e)}/{R_{\rm MC}(\mu/e)}$, is less than 1 is the atmospheric neutrino's anomaly. In Table \ref{table2}, sub-GeV experiments detect the visible-energy less than 1.33GeV, and in the second column (total exposure), left number represents the detector exposure in which fully contained events are detected and right numbers partially contained events. 
\par
The ratios ${R_{\rm expt}(\mu/e)}$ and ${R_{\rm MC}(\mu/e)}$ are defined as
\begin{mathletters}
\begin{eqnarray}
R_{\rm expt}(\mu/e)&=&\frac{\sum_{\alpha}\int\epsilon_{\mu}(E_{\mu})\sigma_{\mu}(E_{\alpha},E_{\mu})F_{\alpha}(E_{\alpha},\theta_{\alpha})P(\nu_\alpha\to\nu_\mu)dE_{\mu}dE_{\alpha}d\theta_{\alpha}}
        {\sum_{\alpha}\int\epsilon_{e}(E_{e})\sigma_{e}(E_{\alpha},E_{e})F_{\alpha}(E_{\alpha},\theta_{\alpha})P(\nu_\alpha\to\nu_e)dE_{e}dE_{\alpha}d\theta_{\alpha}}, \\
R_{\rm MC}(\mu/e)&=&\frac{\sum_{\alpha}\int\epsilon_{\mu}(E_{\mu})\sigma_{\mu}(E_{\alpha},E_{\mu})F_{\alpha}(E_{\alpha},\theta_{\alpha})dE_{\mu}dE_{\alpha}d\theta_{\alpha}}
          {\sum_{\alpha}\int\epsilon_{e}(E_{e})\sigma_{e}(E_{\alpha},E_{e})F_{\alpha}(E_{\alpha},\theta_{\alpha})dE_{e}dE_{\alpha}d\theta_{\alpha}},
\end{eqnarray}
\end{mathletters}
where the summation $\sum_\alpha$ are taken in $\mu$, e neutrino and $\mu,\ e$  untineutrino. $\epsilon_{\beta}(E_{\beta})$ is the detection efficiency of the detector for $\beta$-type charged lepton with energy $E_{\beta}$, $\sigma_{\beta}$ is the differential cross section of $\nu_{\beta}$ and $F_{\alpha}(E_{\alpha},\theta_{\alpha})$ is the incident $\nu_{\alpha}$ flux with energy $E_{\alpha}$ and zenith angle $\theta_{\alpha}$. $P(\nu_\alpha\to\nu_\beta)$ is the transition probability Eq.~(10) and it depends on the energy $E_\alpha$ and the distance $L$ which depends on zenith angle $\theta_\alpha$ as $L=\sqrt{(r+h)^2-r^2\sin^2\theta_\alpha}-r\cos\theta_\alpha$, where $r$ is the radius of the Earth and $h$ is the altitude of production point of atmospheric neutrino.
\par
Although informations of $F_\alpha(E_\alpha,\theta_\alpha)$ etc. are given in Refs. \cite{GAISSER,HONDA,AGRAWAL}, we use the MC predictions for $f_\alpha(E_\alpha, \theta_\alpha)\equiv\sum_{\alpha}\int\epsilon_{\mu}(E_{\mu})\sigma_{\mu}(E_{\alpha},E_{\mu})F_{\alpha}(E_{\alpha},\theta_{\alpha})dE_{\mu}$ in Ref. \cite{HIRATA} for sub-GeV experiment and Ref. \cite{FUKUDA} for multi-GeV experiment. Explicit $E_\alpha$ dependence of $f_\alpha(E_\alpha,\theta_\alpha)$ are shown in Appendix B. Since $P(\nu_\alpha\to\nu_\mu)$ and $P(\nu_\alpha\to\nu_e)$ are the functions of $(\Delta m_{12}^2, \Delta m_{23}^2, \theta_{12}, \theta_{13}, \theta_{23}, L, E)$, the double ratio $R(\mu/e)$ which is integrated in neutrino energy $E$ and zenith angle $\theta$ (related to $L$) is the function of $(\Delta m_{12}^2, \Delta m_{23}^2, \theta_{12}, \theta_{13}, \theta_{23})$. We estimate the $R(\mu/e)$ fixing the parameters $(\delta m_{12}^2, \sin^22\theta)$ on the allowed values  Eq. (15) predicted by the solar neutrino experiments; $\delta m_{12}^2=3\times10^{-5}{\rm eV}^2$, $\sin^22\theta_{12}=0.7$ which corresponds to large angle solution and $\delta m_{12}^2=10^{-5}{\rm eV}^2$, $\sin^22\theta_{12}=0.005$ which corresponds to small angle solution. 
\par
In Fig. \ref{fig4}, we showed the contour plots of double ratio $R(\mu/e)$ on $\tan^2\theta_{13}-\tan^2\theta_{23}$ plane for various $\Delta m^2_{23}$. Contoure lines correspond the upper and lower values of $R(\mu/e)$ in Table II. We showed the plots of sub-GeV experiment in Fig. \ref{fig4}(a)-(d) and plots of multi-GeV one in Fig. \ref{fig4}(e)-(h), and in these figures solid lines denote the large angle solution plots and dotted lines the small angle solution plots. In Fig. (a)-(c), the allowed regions are surrounded by two solid lines (or dotted lines) and in Fig. (d), by two outer lines (or dotted lines) and inner solid line (or dotted line). In Fig. (e)-(h), dotted lines are close in solid lines. In Fig. (e)-(g), allowed regions are surrounded by two outer solid lines (or dotted lines) and inner solid line (or dotted line) and in Fig. (h), by two solid lines (or dotted lines).   
\par
Observing the allowed regions obtained terrestrial neutrino experiments Fig. \ref{fig3} and the present allowed regions obtained atmospheric neutrino experiments, we obtain the allowed regions satisfying all experiments. First, we show the allowed regions of $(\theta_{13},\ \theta_{23})$ for the large angle solution and small angle solution in sub-GeV experiment:
\begin{mathletters}
\begin{eqnarray}
& \qquad\qquad{\rm large\ angle\ solution} & {\rm small\ angle\ solution} \nonumber\\
&{\rm for}\ \ \Delta m_{23}^2=20{\rm eV}^2 \ \ \ {\rm no\ allowed\ region}\qquad\qquad\quad\ \ \ & {\rm no\ allowed\ region} \\
&{\rm for}\ \ \Delta m_{23}^2=2{\rm eV}^2 \ \ \ \  
          (\theta_{13}<4^\circ,\ 18^\circ<\theta_{23}<26^\circ)\quad\ \ \ 
          &
          \left\{\begin{array}{l}
          (\theta_{13}<4^\circ,\ 24^\circ<\theta_{23}<26^\circ) \\
          (\theta_{13}<2^\circ,\ 64^\circ<\theta_{23}<66^\circ)
          \end{array}\right. \\
&{\rm for}\ \ \Delta m_{23}^2=0.2{\rm eV}^2 \ \   
          (\theta_{13}<4^\circ,\ 18^\circ<\theta_{23}<62^\circ)\ \ \ \ \ \ &
          (\theta_{13}<4^\circ,\ 24^\circ<\theta_{23}<66^\circ)  \\
&{\rm for}\  \Delta m_{23}^2=0.02{\rm eV}^2 \  
          \left\{\begin{array}{l}
          (\theta_{13}<12^\circ,\ 17^\circ<\theta_{23}<38^\circ) \\
          (\theta_{13}<12^\circ,\ 47^\circ<\theta_{23}<63^\circ)
          \end{array}\right. &
          \left\{\begin{array}{l}
          (\theta_{13}<14^\circ,\ 23^\circ<\theta_{23}<41^\circ) \\
          (\theta_{13}<14^\circ,\ 49^\circ<\theta_{23}<67^\circ)
          \end{array}\right. \\
&{\rm for}\ \Delta m_{23}^2=0.002{\rm eV}^2 
          \left\{\begin{array}{l}
          (\theta_{13}<29^\circ,\ 27^\circ<\theta_{23}<52^\circ) \\
          (39^\circ<\theta_{13}<59^\circ,\ 76^\circ<\theta_{23})
          \end{array}\right. &
          \left\{\begin{array}{l}
          (\theta_{13}<34^\circ,\ 34^\circ<\theta_{23}<56^\circ) \\
          (34^\circ<\theta_{13}<56^\circ,\ 56^\circ<\theta_{23})
          \end{array}\right. 
\end{eqnarray}
\end{mathletters}
second, in multi-GeV experiment:
\begin{mathletters}
\begin{eqnarray}
& \qquad\qquad\qquad{\rm large\ angle\ solution} & {\rm small\ angle\ solution} \nonumber\\
&{\rm for}\ \ \Delta m_{23}^2=20{\rm eV}^2\qquad {\rm no\ allowed\ region}\qquad\qquad\ \  & {\rm no\ allowed\ region} \\
&{\rm for}\ \ \Delta m_{23}^2=2{\rm eV}^2 \ \ \ \ 
          \left\{\begin{array}{l}
          (\theta_{13}<4^\circ,\ 24^\circ<\theta_{23}<26^\circ) \\
          (\theta_{13}<2^\circ,\ 64^\circ<\theta_{23}<66^\circ)
          \end{array}\right. &
          \left\{\begin{array}{l}
          (\theta_{13}<4^\circ,\ 24^\circ<\theta_{23}<26^\circ) \\
          (\theta_{13}<2^\circ,\ 64^\circ<\theta_{23}<66^\circ)
          \end{array}\right. \\
&{\rm for}\ \ \Delta m_{23}^2=0.2{\rm eV}^2 \ \ \  
          (\theta_{13}<4^\circ,\ 24^\circ<\theta_{23}<66^\circ)\ \ \  &
          (\theta_{13}<4^\circ,\ 24^\circ<\theta_{23}<66^\circ) \\
&{\rm for}\ \ \Delta m_{23}^2=0.02{\rm eV}^2 \ \  
          (\theta_{13}<12^\circ,\ 41^\circ<\theta_{23}<49^\circ)\ \  &
          (\theta_{13}<12^\circ,\ 41^\circ<\theta_{23}<49^\circ) \\
&{\rm for}\ \ \Delta m_{23}^2=0.002{\rm eV}^2 \  
          (22^\circ<\theta_{13}<68^\circ,\ 45^\circ<\theta_{23}) &
          (22^\circ<\theta_{13}<68^\circ,\ 45^\circ<\theta_{23})
\end{eqnarray}
\end{mathletters}
\par
If we combine the LSND experiment with above terrestrial data, allowed regions are restricted as follows,\\
sub-GeV case:
\begin{mathletters}
\begin{eqnarray}
& \qquad\qquad\qquad{\rm large\ angle\ solution}\ \ \ \  & {\rm small\ angle\ solution} \nonumber\\
&{\rm for}\ \ \Delta m_{23}^2=2{\rm eV}^2 \qquad   
          (2^\circ<\theta_{13}<4^\circ, 18^\circ<\theta_{23}<26^\circ)
          &
          \left\{\begin{array}{l}
          (2^\circ<\theta_{13}<4^\circ,\\
          \qquad 24^\circ<\theta_{23}<26^\circ) \\
          (0.8^\circ<\theta_{13}<2^\circ,\\
          \qquad 64^\circ<\theta_{23}<66^\circ)
          \end{array}\right. \\
& {\rm for}\ \ \Delta m_{23}^2=0.2{\rm eV}^2 \ \ \ \   
          {\rm no\ allowed\ region}\qquad\qquad\qquad 
          &{\rm no\ allowed\ region}\qquad                  
\end{eqnarray}
\end{mathletters}
multi-GeV case:
\begin{mathletters}
\begin{eqnarray}
& \qquad\qquad\qquad{\rm large\ angle\ solution}\ \ \ \  & {\rm small\ angle\ solution} \nonumber\\
&{\rm for}\ \ \Delta m_{23}^2=2{\rm eV}^2 \ \ \ 
          \left\{\begin{array}{l}
          (2^\circ<\theta_{13}<4^\circ,\\
          \qquad\ 24^\circ<\theta_{23}<26^\circ) \\
          (0.8^\circ<\theta_{13}<2^\circ,\\
          \qquad 64^\circ<\theta_{23}<66^\circ)
          \end{array}\right. &
          \left\{\begin{array}{l}
          (2^\circ<\theta_{13}<4^\circ,\\
          \qquad 24^\circ<\theta_{23}<26^\circ) \\
          (0.8^\circ<\theta_{13}<2^\circ,\\
          \qquad 64^\circ<\theta_{23}<66^\circ)
          \end{array}\right. \\
& {\rm for}\ \ \Delta m_{23}^2=0.2{\rm eV}^2 \ \    
          {\rm no\ allowed\ region}\qquad\ \ \  
          &{\rm no\ allowed\ region}
\end{eqnarray}
\end{mathletters}
\par
Although there are allowed regions satisfying both terrestrial and atmospheric experimental restrictions, all of these solutions do not satisfy the zenith angle dependence of $R(\mu/e)$ for atmospheric neutrino experiments. From resent SuperKamiokande experiments \cite{SUPERKAMIOKANDE,KANEYUKI}, zenith angle dependence seems to be more definite than that obtained previously \cite{HIRATA,FUKUDA}: the double ratio $R(\mu/e)$ for sub-GeV experiment seems to decrease monotonically as zenith angle $\theta$ increases from $\theta=0^\circ$ to $\theta=180^\circ$ and $R(\mu/e)$ for multi-GeV experiments seems to decrease as zenith angle $\theta$ increases from $\theta=0^\circ$ to $\theta=180^\circ$.  Among the allowed solutions obtained above, Eqs. (21) and (22) (or (23) and (24)), the large angle  ($\sin^22\theta_{12}\sim0.7$) solutions with $\theta_{23}<45^\circ$ satisfy the monotonic decreasing of the $R(\mu/e)$ in sub-GeV neutrino experiment. This solution also decreases in multi-GeV experiment as zenith angle increases. 
\par
In Fig. \ref{fig5}(a), we showed the zenith angle ($\cos\theta$) dependence of $R(\mu/e)$ in sub-GeV experiment on the typical parameters for large angle solution $(\Delta m_{12}^2=3\times10^{-5}{\rm eV}^2,\ \sin^22\theta_{12}=0.7,\ \Delta m_{23}=0.2{\rm eV}^2,\ \theta_{13}=4^\circ,\ \theta_{23}=25^\circ)$ by solid curve and small angle solution $(\Delta m_{12}^2=10^{-5}{\rm eV}^2,\ \sin^22\theta_{12}=0.007,\ \Delta m_{23}=0.2{\rm eV}^2,\ \theta_{13}=4^\circ,\ \theta_{23}=25^\circ)$ by dotted curve. Experimental data is referred to SuperKamiokande \cite{KANEYUKI}. Fig. \ref{fig5}(b) represents the zenith angle ($\cos\theta$) dependence of $R(\mu/e)$ in multi-GeV experiment on the parameters for large angle solution $(\Delta m_{12}^2=3\times10^{-5}{\rm eV}^2,\ \sin^22\theta_{12}=0.7,\ \Delta m_{23}=0.2{\rm eV}^2,\ \theta_{13}=4^\circ,\ \theta_{23}=30^\circ)$ (solid curve) and small angle solution $(\Delta m_{12}^2=10^{-5}{\rm eV}^2,\ \sin^22\theta_{12}=0.007,\ \Delta m_{23}=0.2{\rm eV}^2,\ \theta_{13}=4^\circ,\ \theta_{23}=30^\circ)$ (dotted curve).
\par
We summarize the solution satisfying the solar, terrestrial and atmospheric experiments as follows;
\begin{mathletters}
\begin{eqnarray}
{\rm for}&&\ \ \Delta m_{23}^2=2{\rm eV}^2\nonumber\\
&& \left\{\begin{array}{l}
   \Delta m_{12}^2=4\times10^{-6} - 7\times10^{-5}{\rm eV}^2,\ \ \sin^22\theta_{12} = 0.6 - 0.9,\\
   \theta_{13}<4^\circ,\ \ 24^\circ<\theta_{23}<26^\circ,
   \end{array}\right.\\
{\rm for}&&\ \ \Delta m_{23}^2=0.2{\rm eV}^2\nonumber\\   
&& \left\{\begin{array}{l}
   \Delta m_{12}^2=4\times10^{-6} - 7\times10^{-5}{\rm eV}^2,\ \ \sin^22\theta_{12} = 0.6 - 0.9,\\
   \theta_{13}<4^\circ,\ \ 24^\circ<\theta_{23}<45^\circ.
   \end{array}\right.\\
{\rm for}&&\ \ \Delta m_{23}^2=0.02{\rm eV}^2\nonumber\\   
&& \left\{\begin{array}{l}
   \Delta m_{12}^2=4\times10^{-6} - 7\times10^{-5}{\rm eV}^2,\ \ \sin^22\theta_{12} = 0.6 - 0.9,\\
   \theta_{13}<12^\circ,\ \ \theta_{23}\sim40^\circ.
   \end{array}\right.   
\end{eqnarray}
\end{mathletters}
If we include the LSND experiment in terrestrial ones, solutions, Eqs. (25a) and (25b), are favoured.  
\section{Discussions}
We analyzed the solar, terrestrial and atmospheric neutrino experiments using the three-flavor neutrinos framework and got the allowed regions of the parameters $(\Delta m_{12}^2,\ \sin^22\theta_{12},\ \Delta m_{23}^2,\ \theta_{13},\ \theta_{23})$. In solar neutrino experiments, we got the large angle solution $(\Delta m_{12}^2,\ \sin^22\theta_{12})=(4\times10^{-6} - 7\times10^{-5}{\rm eV}^2,\ 0.6 - 0.9)$ and small angle solution $(3\times10^{-6} - 1.2\times10^{-5}{\rm eV}^2,\ 0.003 - 0.01)$ for $\theta_{13}=0^\circ - 20^\circ$. When $\theta_{13}$ increases from $25^\circ$ to $50^\circ$, large angle and small angle solutions merge. From the terrestrial and atmospheric neutrino experiments (not considering the zenith angle dependence), we got the allowed regions:  $(\theta_{13}<4^\circ,\ 24^\circ<\theta_{23}<26^\circ)$ for large angle solution and  $\Delta m_{23}^2=2{\rm eV}^2$, $(\theta_{13}<4^\circ,\ 24^\circ<\theta_{23}<26^\circ)$ and $(\theta_{13}<2^\circ,\ 64^\circ<\theta_{23}<66^\circ)$ for small angle solution and $\Delta m_{23}^2=2{\rm eV}^2$, $(\theta_{13}<4^\circ,\ 24^\circ<\theta_{23}<62^\circ)$ for large angle solution and $\Delta m_{23}^2=0.2{\rm eV}^2$, $(\theta_{13}<4^\circ,\ 24^\circ<\theta_{23}<66^\circ)$ for small angle solution and $\Delta m_{23}^2=0.2{\rm eV}^2$, $(\theta_{13}<12^\circ,\ \theta_{23}\sim40^\circ,\ 48^\circ)$ for large angle solution and $\Delta m_{23}^2=0.02{\rm eV}^2$, $(\theta_{13}<14^\circ,\ \theta_{23}\sim41^\circ,\ 49^\circ)$ for small angle solution and $\Delta m_{23}^2=0.02{\rm eV}^2$.  When the zenith angle dependences are considered, the allowed regions are restricted to $(\Delta m_{12}^2=4\times10^{-6}-7\times10^{-5}{\rm eV}^2,\ \sin^22\theta_{12}=0.6-0.9,\ \theta_{13}<4^\circ,\ 24^\circ<\theta_{23}<26^\circ)$ for $\Delta m_{23}^2=2{\rm eV}^2$, $(\Delta m_{12}^2=4\times10^{-6}-7\times10^{-5}{\rm eV}^2,\ \sin^22\theta_{12}=0.6-0.9,\ \theta_{13}<4^\circ,\ 24^\circ<\theta_{23}<45^\circ)$ for $\Delta m_{23}^2=0.2{\rm eV}^2$ and $(\Delta m_{12}^2=4\times10^{-6}-7\times10^{-5}{\rm eV}^2,\ \sin^22\theta_{12}=0.6-0.9,\ \theta_{13}<12^\circ,\ \theta_{23}\sim 40^\circ)$ for $\Delta m_{23}^2=0.02{\rm eV}^2$. If we include LSND experiment in terrestrial experiments, allowed regions are restricted to $(\Delta m_{12}^2=4\times10^{-6}-7\times10^{-5}{\rm eV}^2,\ \sin^22\theta_{12}=0.6-0.9,\ 2^\circ<\theta_{13}<4^\circ,\ 24^\circ<\theta_{23}<26^\circ)$ and $\Delta m_{23}^2=2-0.2{\rm eV}^2$.
\par
Finally, we present the neutrino mixing matrix Eq. (2) numerically for the allowed solutions Eqs. (25a), (25b) and (25c);
\begin{mathletters}
\begin{eqnarray}
{\rm for}&&\ \ \Delta m_{23}^2=2{\rm eV}^2\nonumber\\
&&U=\left(\begin{array}{ccc}
       0.81 \leftrightarrow 0.90 & 0.43 \leftrightarrow 0.58 & 0.0 \leftrightarrow 0.07\\
       -0.39 \leftrightarrow -0.56 & 0.71 \leftrightarrow 0.83 & 0.41 \leftrightarrow 0.44\\
       0.12 \leftrightarrow 0.26 & -0.33 \leftrightarrow -0.42 & 0.90 \leftrightarrow 0.91
       \end{array}\right),\\
{\rm for}&&\ \ \Delta m_{23}^2=0.2{\rm eV}^2\nonumber\\
&&U=\left(\begin{array}{ccc}
       0.81 \leftrightarrow 0.90 & 0.43 \leftrightarrow 0.58 & 0.0 \leftrightarrow 0.07\\
       -0.30 \leftrightarrow -0.56 & 0.54 \leftrightarrow 0.83 & 0.41 \leftrightarrow 0.71\\
       0.12 \leftrightarrow 0.41 & -0.33 \leftrightarrow -0.64 & 0.71 \leftrightarrow 0.91
       \end{array}\right),\\
{\rm for}&&\ \ \Delta m_{23}^2=0.02{\rm eV}^2\nonumber\\
&&U=\left(\begin{array}{ccc}
       0.79 \leftrightarrow 0.90 & 0.42 \leftrightarrow 0.58 & 0.0 \leftrightarrow 0.24\\
       -0.33 \leftrightarrow -0.57 & 0.53 \leftrightarrow 0.69 & 0.62 \leftrightarrow 0.64\\
       0.11 \leftrightarrow 0.38 & -0.52 \leftrightarrow -0.66 & 0.74 \leftrightarrow 0.77
       \end{array}\right).       
\end{eqnarray}
\end{mathletters}             

\appendix
\section{Numerical analysis for $R$ in solar neutrino}
The ratio $R$ of the detected $e$ neutrino flux to the expected $e$ neutrino flux deduced from SSM is expressed as
\begin{equation}
R=\frac{\int^{E_{max}}_{E_{min}} P^{\rm MSW}_{3\nu}(E)f(E)dE}{\int^{E_{max}}_{E_{min}} f(E) dE},
\end{equation}
where $E$ is the neutrino energy and $f(E)$ is the summation of products of $f_i(E)$ and $c^D_i$. 
\begin{equation}
f(E)=\sum_{i}c_i^Df_i(E).
\end{equation}
$f_i(E)$ represents the neutrino flux produced in the reaction of type $i$, where $i=pp,\ ^{13}{\rm N},\ ^{15}{\rm O},\ ^7{\rm Be},\ ^8{\rm B}$ and $pep$, at the center of the sun. $c^D_i$ is the detector sensitivity of detector $D$ for the neutrino flux of type $i$. From the SSM \cite{BAHCALL,PDG96}, neutrino fluxes at 1AU are expressed as the function of $E$ (numerical number without dimension of neutrino energy measured in the unit MeV) as
\begin{mathletters}
\begin{eqnarray}
f_{pp}(E)&=&E^{1.88}(4.358-0.0946E-24.48E^2)\times10^{12},\ \ \ (E<0.420{\rm MeV})\\
f_{Be1}(E)&=&8\times10^{8},\ \ \ (0.384-0.25{\rm MeV}< E<0.384+0.25{\rm MeV})\\
f_{N}(E)&=&E^{1.88}(4.194-4.541E+0.870E^2)\times10^{9},\ \ \ (E<1.199{\rm MeV})\\
f_{Be2}(E)&=&4\times10^{9},\ \ \ (0.862-0.5{\rm MeV}< E<0.862+0.5{\rm MeV})\\ 
f_{O}(E)&=&E^{1.88}(12.74-7.919E+0.328E^2)\times10^{8},\ \ \ (E<1.732{\rm MeV})\\ 
f_{pep}(E)&=&1.5\times10^{8},\ \ \ (1.442-0.5{\rm MeV}< E<1.442+0.5{\rm MeV})\\
f_{B}(E)&=&E^{1.88}(81211-11500E-407.1E^2).\ \ \ (E<14.02{\rm MeV}) 
\end{eqnarray}
\end{mathletters}
We estimate the detector sensitivity which is normalized as $c_{B}^D=1$ using the Bahcall's result (Table \ref{table1}) \cite{BAHCALL},
\begin{mathletters}
\begin{eqnarray}
&&{\rm Ga\ experiment} : c_{pp}^{Ga}=9.62\times10^{-4},\ c_{N}^{Ga}=2.73\times10^{-3},\ c_{Be1,Be2}^{Ga}=3.61\times10^{-3}, \nonumber \\
&& \qquad\qquad\qquad c_{O}^{Ga}=4.30\times10^{-3},\ c_{pep}^{Ga}=8.93\times10^{-3},\ c_{B}^{Ga}=1.0,\\
&&{\rm Cl\ experiment} : c_{pp}^{Cl}=0,\ c_{N}^{Cl}=4.82\times10^{-4},\ c_{Be1}^{Cl}=0,\ c_{Be2}^{Cl}=5.08\times10^{-4}, \nonumber\\
&&\qquad\qquad\qquad c_{O}^{Cl}=7.64\times10^{-4},\ c_{pep}^{Cl}=1.44\times10^{-3},\ c_{B}^{Cl}=1.0,\\
&&{\rm Kamiokande\ experiment} : c_{pp}^{Kam}=0,\ c_{N}^{Kam}=0,\ c_{Be1}^{Kam}=0,\ c_{Be2}^{Kam}=0,\ c_{O}^{Kam}=0, \nonumber \\
&&\qquad\qquad\qquad c_{pep}^{Kam}=0,\ c_{B}^{Kam}=1.0.
\end{eqnarray}
\end{mathletters}
Numerical values used for parameters at the center of the sun are as follows:
\begin{mathletters}
\begin{eqnarray}
&&A=2\sqrt{2}G_FN_eE=2\sqrt{2}G_F(Y_e/m_n)\rho E,\\
&&\qquad G_F=1.17\times10^{-23}{\rm eV}^{-2}\ \ {\rm (Fermi\ constant)},\nonumber\\
&&\qquad Y_e=1/2\ \ {\rm (the\ number\ of\ electrons\ per\ nucleon)},\nonumber\\
&&\qquad m_n=939\,{\rm MeV}\ \ {\rm (the\ nucleon\ mass)},\ \ \ \rho=156\,{\rm g/{cm}}^3\ \ {\rm (the\ density)},\nonumber\\
&&\left|\frac1{N_e}\frac{dN_e}{dx}\right|_0=2.91\times10^{-15}. 
\end{eqnarray}
\end{mathletters}

\section{Explicit $E_\alpha$ dependence of $\lowercase{f}_\alpha(E_\alpha,\theta_\alpha)$}
 We take the $E_\alpha$ dependence of $f_\alpha(E_\alpha,\theta_\alpha)$ from the results estimated in Monte Callro calculation. We neglect the $\theta_\alpha$ dependence of $f_\alpha(E_\alpha,\theta_\alpha)$. For the sub-Gev experiments, we take it from Ref. \cite{HIRATA} as 
\begin{mathletters}
\begin{eqnarray}
f_{\nu_\mu}(E, \theta)&=&N_s(-2.89E^{-2.55}+22.9E^{-1.55}-12.7E^{-0.55}),\\
f_{\nu_e}(E, \theta)&=&N_s(-1.18E^{-2.55}+9.98E^{-1.55}-3.90E^{-0.55}),
\end{eqnarray}
\end{mathletters}
and for the multi-GeV experiments from Ref. \cite{FUKUDA} as 
\begin{mathletters}
\begin{eqnarray}
f_{\nu_\mu}(E, \theta)&=&N_m(42.1-383E^{-2}+374E^{-1}-2.42E+0.0403E^2\\
&&-0.000206E^3),\ \ {\rm  fully\mbox{-}and\ partially\mbox{-}contained\ events} \\
f_{\nu_e}(E, \theta)&=&N_m(-9.54-225E^{-2}+249E^{-1}+0.156E-0.000658E^2\\
&&-2.42\times10^{-6}E^3),\ \ {\rm fully\mbox{-}contaiened\ events} 
\end{eqnarray}
\end{mathletters}
where $E$ is the numerical number of neutrino energy measured in unit GeV, and $N_s$ and $N_m$ are constants with the unit: number of events/GeV. 


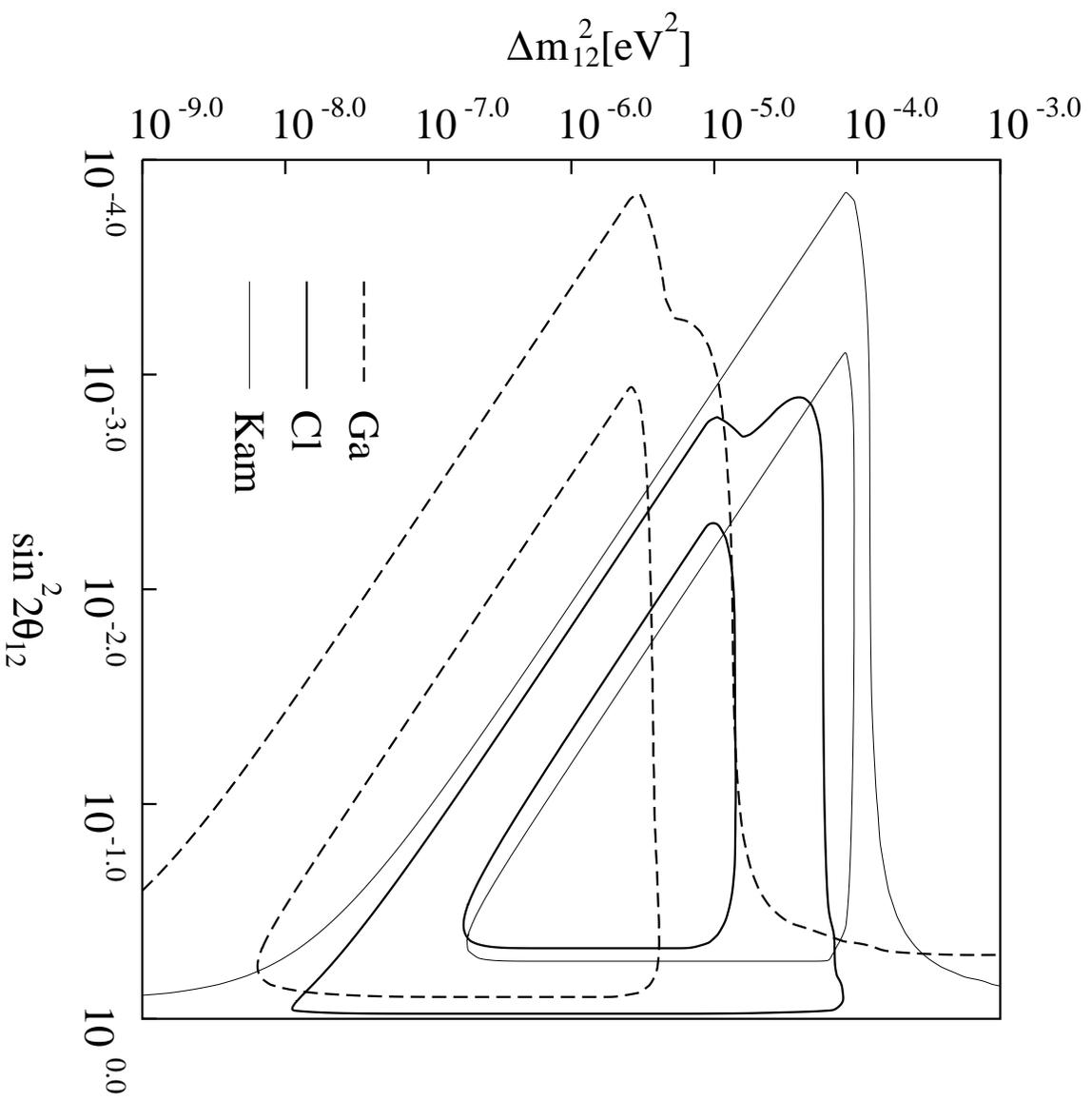
\begin{figure}
\caption{Contour plot of $R$ for $\theta_{13}=0$ on $\sin^22\theta_{12}-\Delta m_{12}^2$ plane. Each two contours denoted as Ga, Cl and Kam corresponds to the upper and lower value of $R$ for Ga, Cl and Kamiokande experiment, respectively. Two solutions are denoted as common areas enclosed by each two contours of Ga, Cl and Kam.}
\label{fig1}
\end{figure}
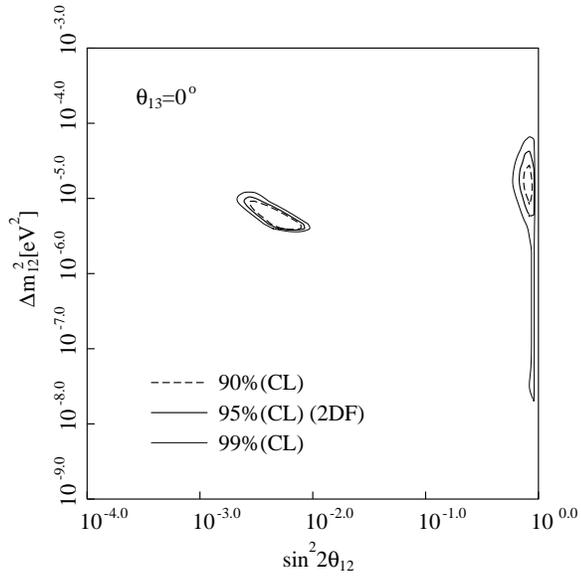
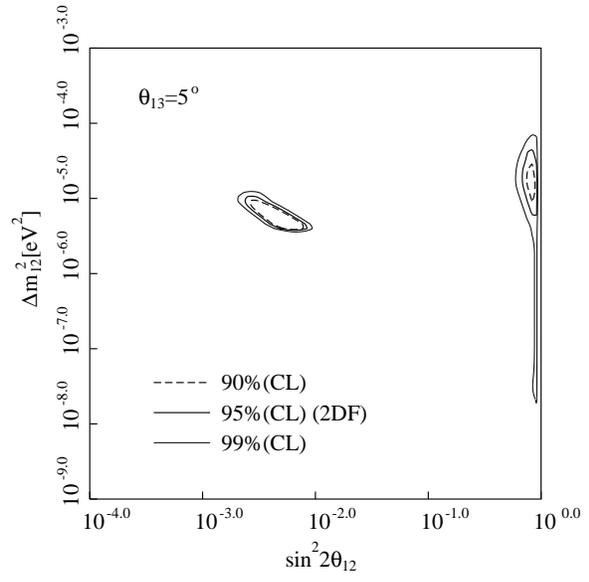
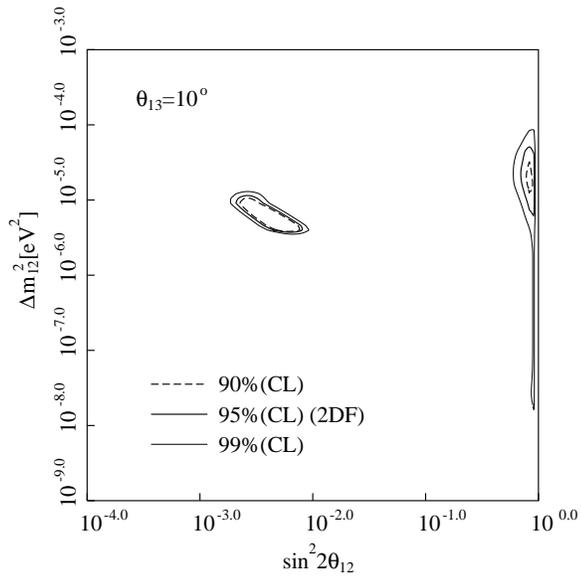
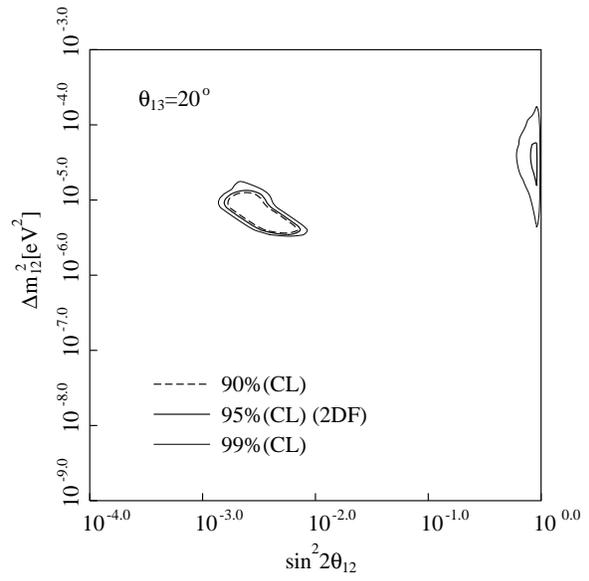
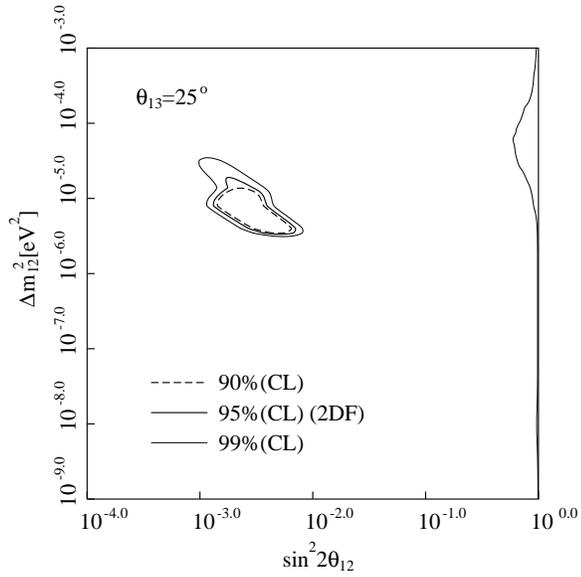
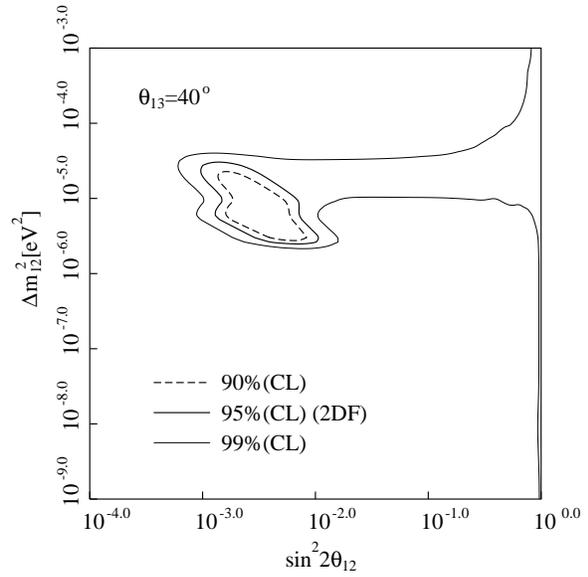
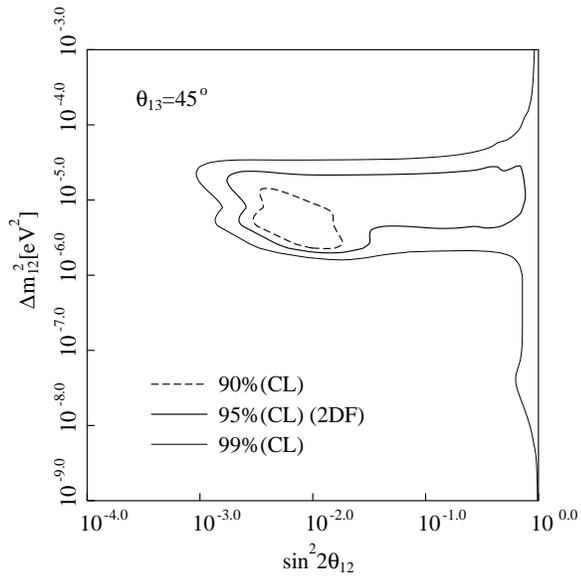
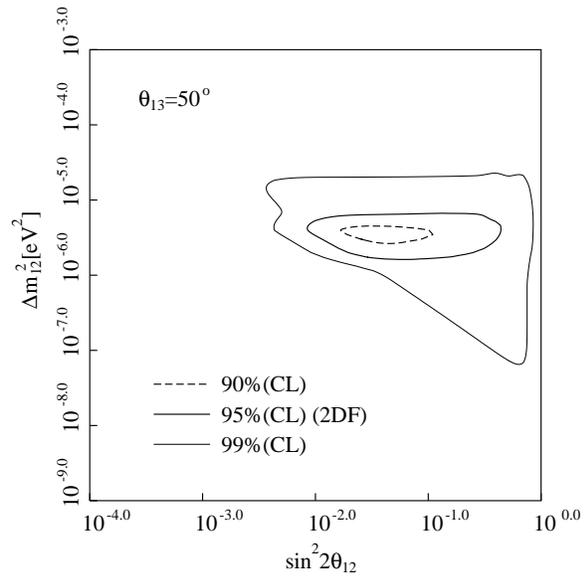
\begin{figure}
\caption{The plots of allowed regions of the combined Ga, Cl and Kam experiments using the $\chi$-square, where the solid thin, solid thick and dotted lines define the regions allowed at $99 \%,\ 95 \%$ and $90 \%$ C.L., respectively. Figs. (a) - (h) show the allowed regions for $\theta_{13}=0^{\circ} - 50^{\circ}$.} 
\label{fig2}
\end{figure}
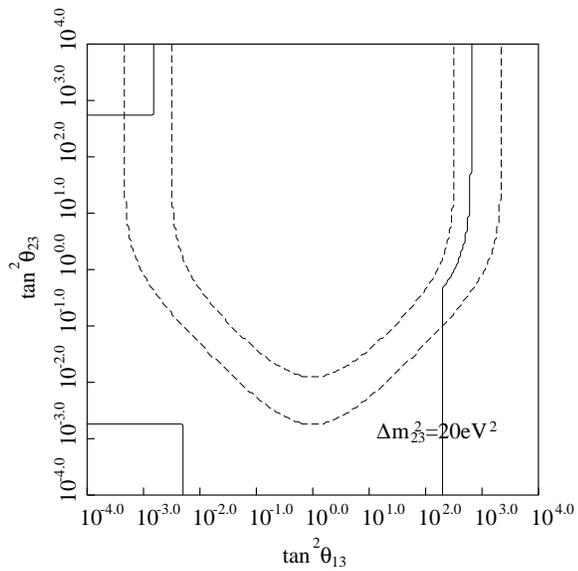
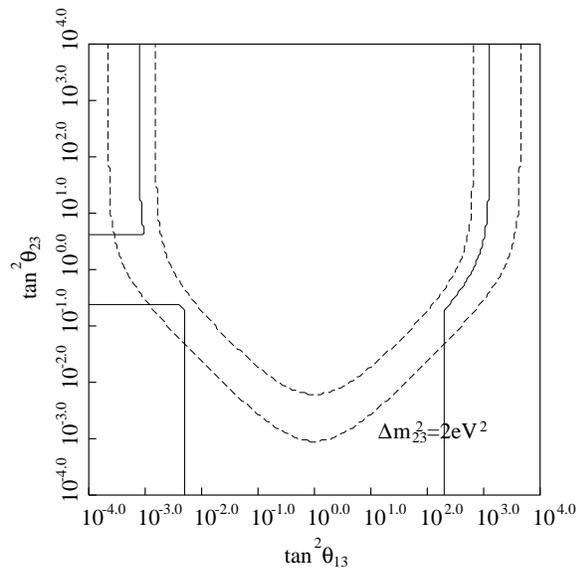
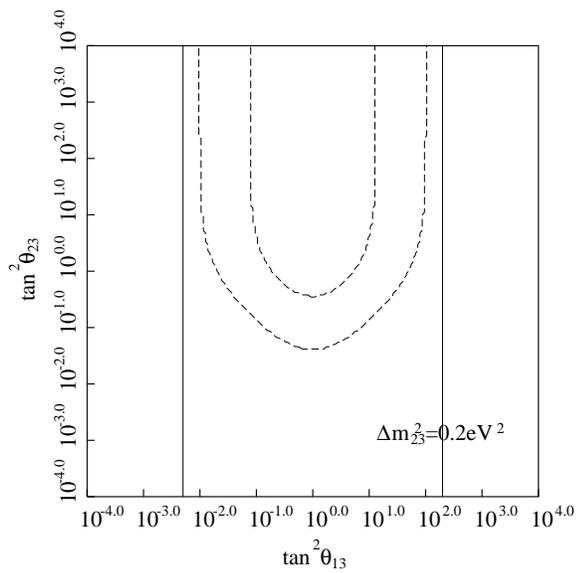
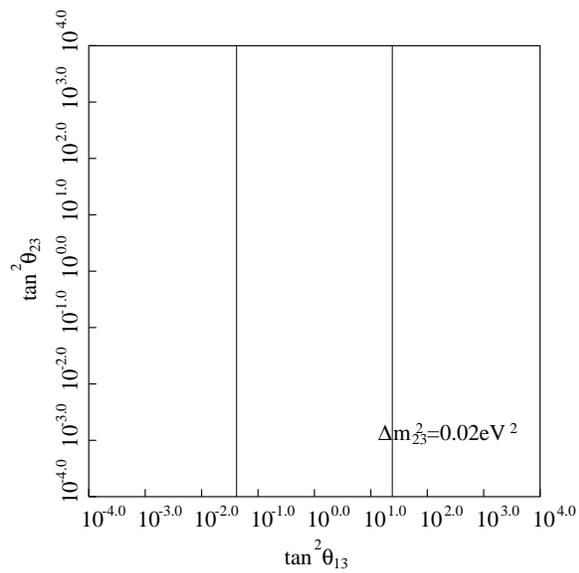
\begin{figure}
\caption{The plots of allowed regions on $\theta_{13}-\theta_{23}$ plane determined by $P$ of terrestrial $\nu_{\mu}\to\nu_{\tau}$, ${\nu}_{\mu}\to\nu_{e}$, $\bar{\nu}_{\mu}\to\bar{\nu}_{e}$ and $\bar{\nu}_e\to\bar{\nu}_e$ experiments. Allowed regions are corners, left and right hand sides surrouned by curves. Curves represent the boundary of $90$ \% C.L. of $P$. $\Delta m_{12}^2$ and $\sin^22\theta_{12}$ are fixed as $10^{-5}{\rm eV}^2$ and 0.8, respectively. $\Delta m_{23}^2$ is fixed to 20eV$^2$(Fig.~3(a)), 2.0eV$^2$(Fig.~3(b)), 0.2eV$^2$(Fig.~(c)) and 0.02eV$^2$(Fig.~(d)). Dotted lines show the allowed regions restricted by the LSND data.} 
\label{fig3}
\end{figure}
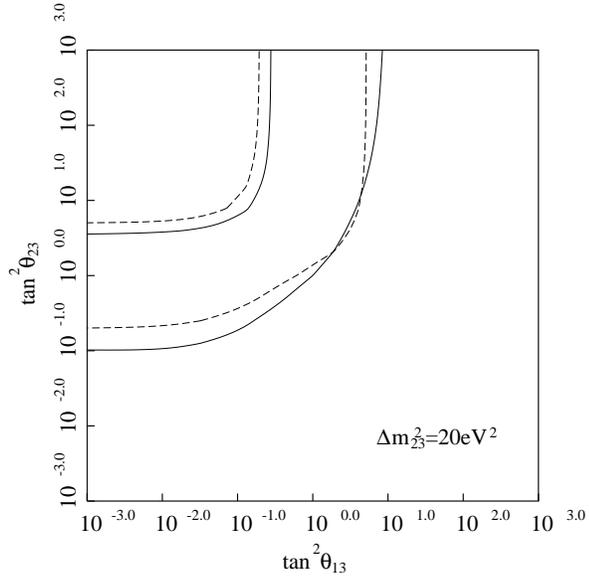
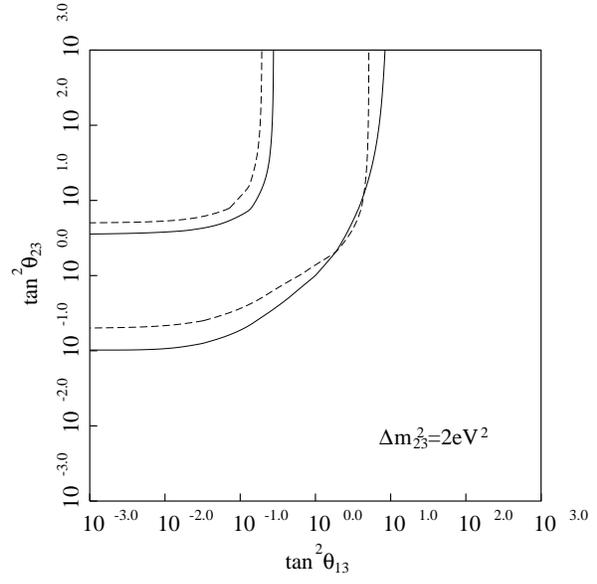
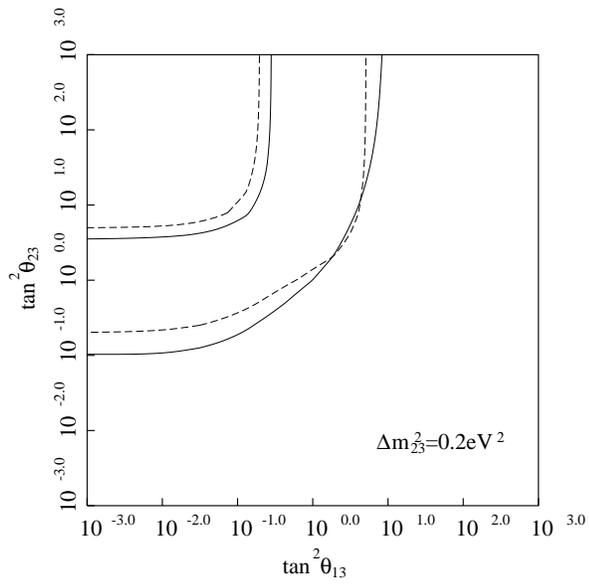
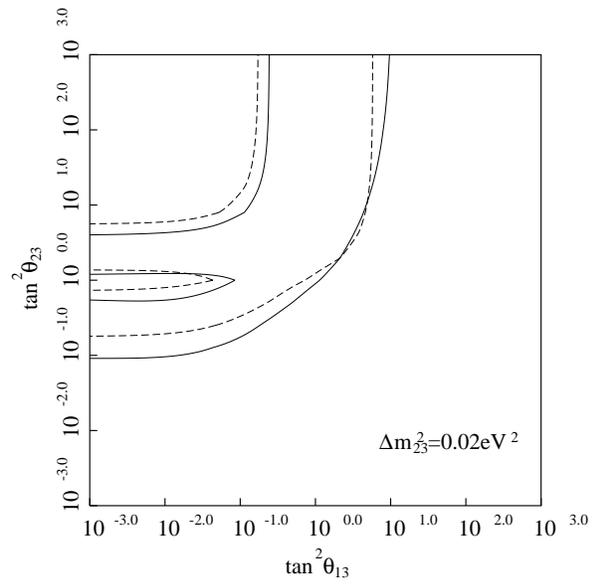
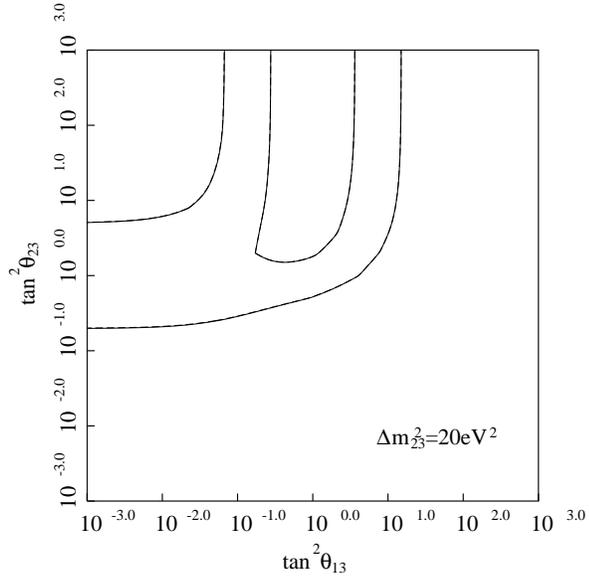
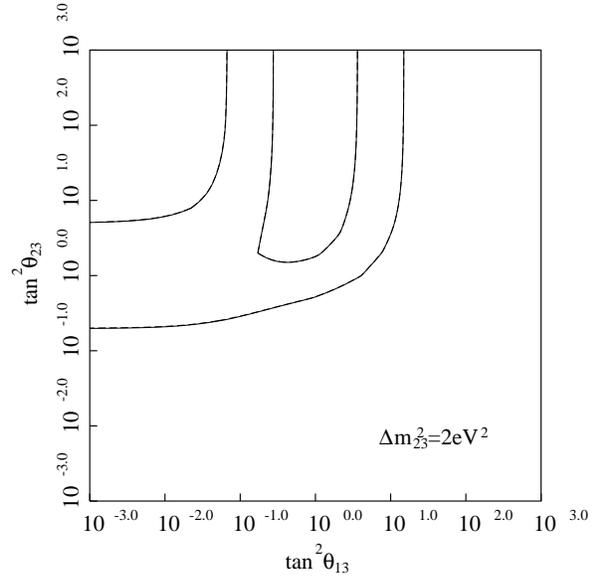
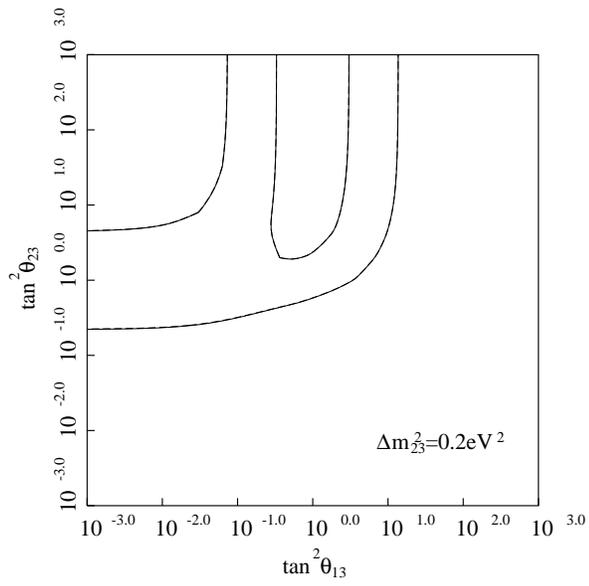
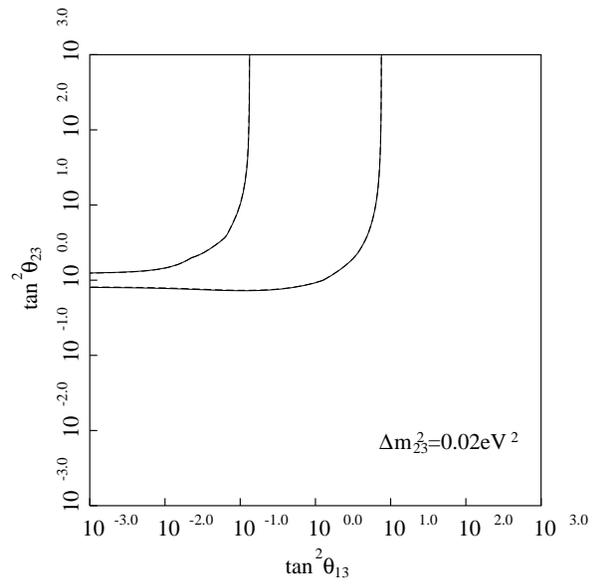
\begin{figure}
\caption{The plots of allowed regions of $R(\mu/e)$ of atmospheric neutrinos for various values of $\Delta m_{23}^2 $ on $\tan^2\theta_{13}-\tan^2\theta_{23}$ .Figs. (a)-(d) show the plots of sub-GeV experiments and Figs. (e)-(h) the ones of multi-GeV experiments. We fix the parameters $(\Delta m_{12}^2$, $\sin^22\theta)$ to be $(3\times10^{-5}{\rm eV}^2,\ 0.7)$ corresponding to the large angle solution (solid lines) and  $(10^{-5}{\rm eV}^2,\ 0.005)$ corresponding to the small angle solution (dotted lines).} 
\label{fig4}
\end{figure}
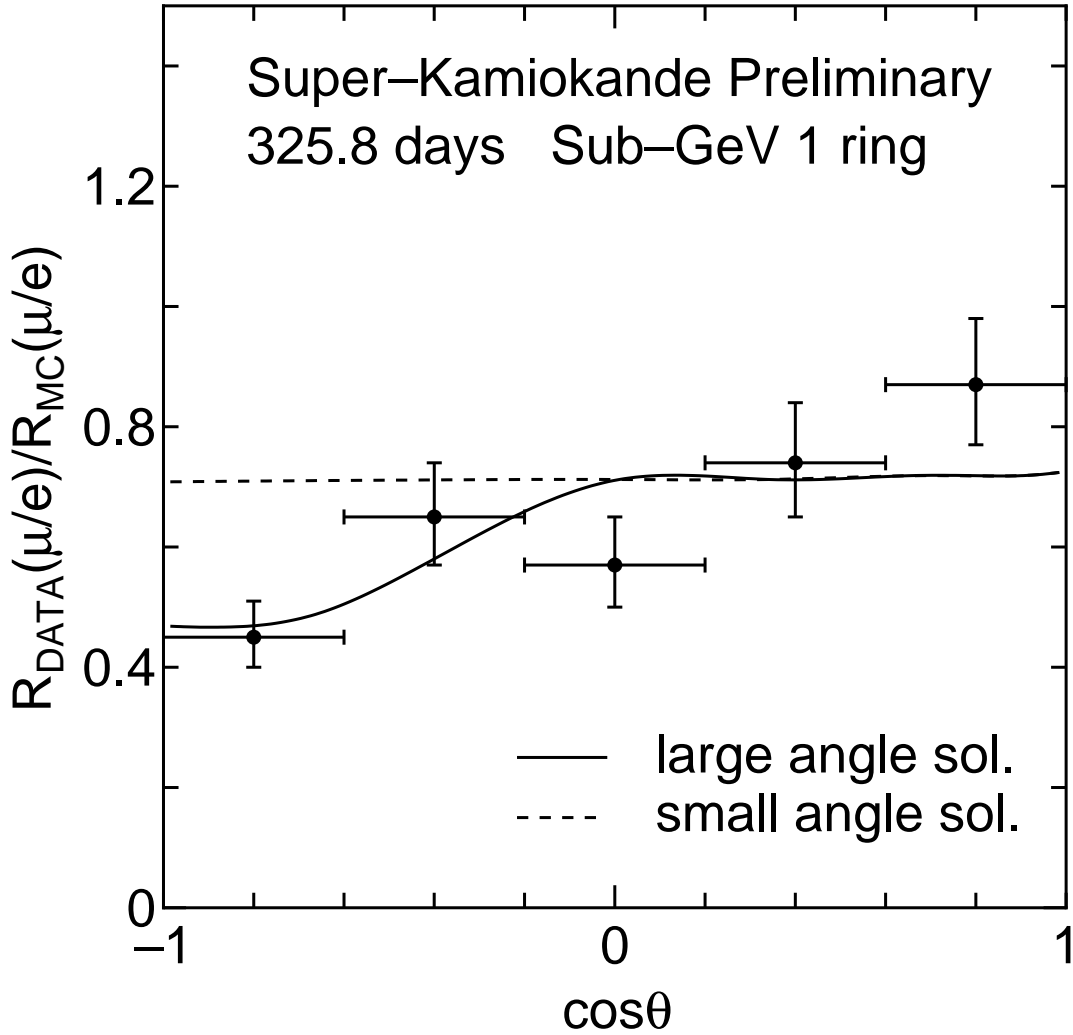
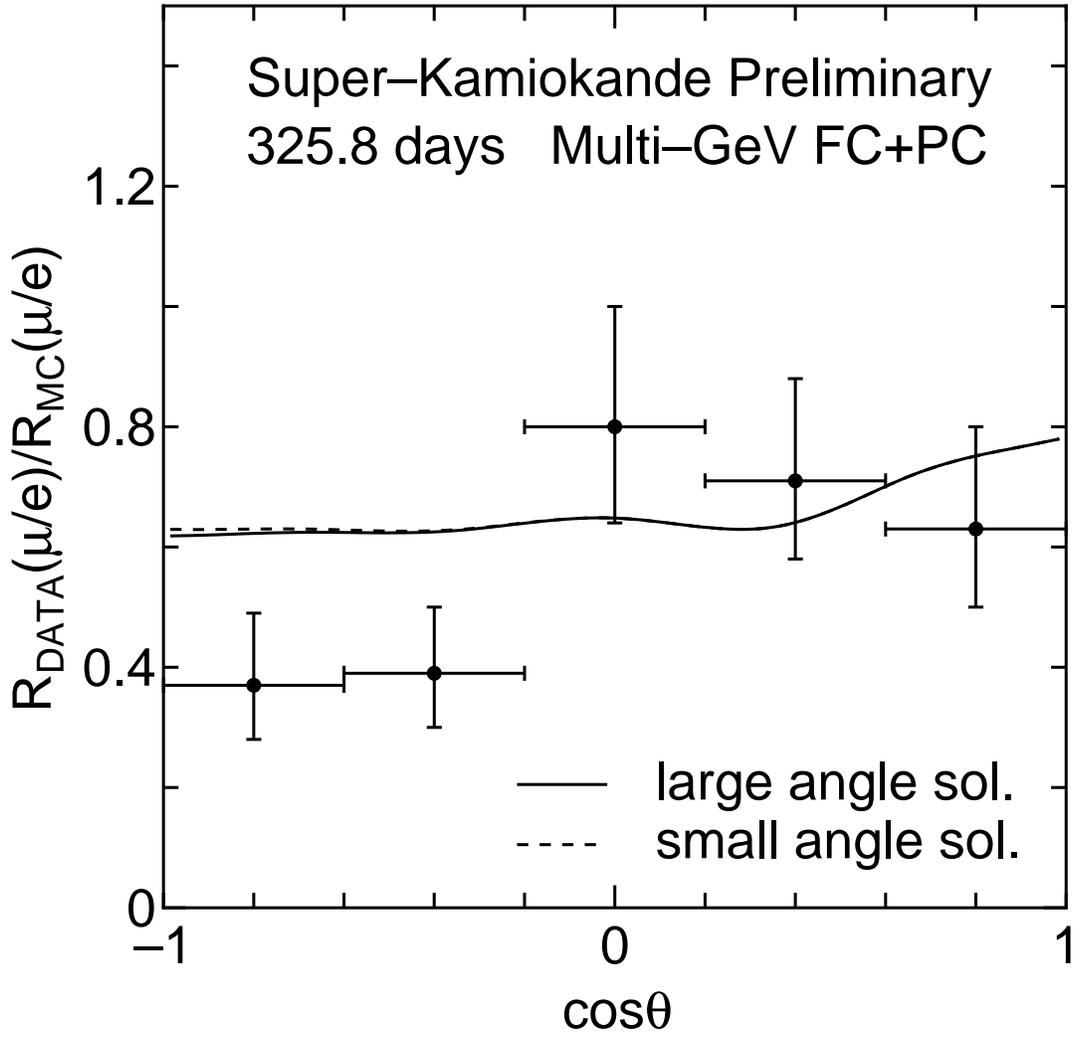
\begin{figure}
\caption{The zenith angle dependence of $R(\mu/e)$ in atmospheric neutrino experiment. Experimental data is referred to Ref. [20]. Fig. (a) represents the sub-GeV experimental case: solid curve is given on a typical parameters for large angle solution $(\Delta m_{12}^2=$ $3\times10^{-5}{\rm eV}^2,$ $\ \sin^22\theta_{12}=0.7,\ \Delta m_{23}=0.2{\rm eV}^2,\ \theta_{13}=4^\circ,\ \theta_{23}=25^\circ)$ and dotted curve on small angle solution $(\Delta m_{12}^2=10^{-5}{\rm eV}^2,\ \sin^22\theta_{12}=0.007,\ \Delta m_{23}=0.2{\rm eV}^2,$ $\ \theta_{13}=4^\circ,$ $\ \theta_{23}=25^\circ)$. Fig. (b) represents the multi-GeV experimental case: solid curve is given on parameters for large angle solution $(\theta_{13}=4^\circ,\ \theta_{23}=30^\circ)$ and dotted curve on small angle solution $(\theta_{13}=4^\circ,$ $\ \theta_{23}=30^\circ)$. }
\label{fig5}
\end{figure}

\begin{table}
\begin{center}
\caption{Individual neutrino contribution (averaged on the cases with and without diffusion) calculated by Bahcall(1995) }.
\label{table1}
\begin{tabular}{ccc}
neutrino source   &  Cl(SNU)  &  Ga(SNU) \\ 
\hline
$pp$              &   0.00    &  70      \\
$pep$             &   0.23    &   3      \\
$^7$Be            &   1.17    &  35      \\
$^8$B             &   6.4     &  14      \\
$^{13}$N          &   0.09    &   3      \\
$^{15}$O          &   0.3     &   5      \\
\hline
total             &   8.2     & 132      \\ 
\end{tabular}
\end{center}
\end{table}

\begin{table}
\begin{center}
\caption{The results of atmospheric experiments predicting the anomaly.  Sub-GeV experiments detect the visible-energy less than 1.33GeV, and in the second column (total exposure), left number represents the detector exposure in which fully contained events are detected and right numbers partially contained events. }
\label{table2}
\begin{tabular}{ccc}
Experiments & Total exposure(ktyr)  &  $R(\mu/e)$(double ratio) \\ 
\hline
Kamiokande(sub-Gev)\cite{HIRATA}   &  4.92  & $0.60\mbox{\small${+0.07\atop-0.06}$}\pm0.05$  \\
Kamiokande(multi-GeV)\cite{FUKUDA}   &  8.2, 6.0  & $0.57\mbox{\small${+0.08\atop-0.07}$}\pm0.07$  \\
IMB \cite{IMB}   &  7.7   & $0.54\pm0.02\pm0.07$  \\
SuperKamiokande(sub-GeV) \cite{KANEYUKI}   &  20, 18  &  $0.63\pm0.03\pm0.05$      \\
SuperKamiokande(multi-GeV) \cite{KANEYUKI}    &  20, 18   &  $0.60\pm0.06\pm0.07$      \\
\end{tabular}
\end{center}
\end{table}

\end{document}